\documentclass[preprint2,twocolumn,times,tighten]{aastex63}

\usepackage{graphicx,times}
\usepackage{subfigure}
\newcommand{\be}{\begin{equation}}
\usepackage{threeparttable}
\usepackage{booktabs}
\newcommand{\ee}{\end{equation}}
\newcommand{\bea}{\begin{eqnarray}}
\newcommand{\eea}{\end{eqnarray}}
\usepackage{color}

\newcommand*{\xu}{\color{magenta}}

\usepackage{enumerate}
\usepackage{amsmath}
\usepackage{cases}
\usepackage{longtable}
\usepackage{hyperref}
\usepackage{epstopdf}
\usepackage{amsmath,bm}
\usepackage{amssymb}
\usepackage{natbib}
\usepackage{morefloats}
\usepackage{multirow}
\usepackage{array}
\usepackage{verbatim}
\usepackage{lineno}

\begin{document}

\title{Mirror acceleration of cosmic rays in a high-$\beta$ medium}

\email{lazarian@astro.wisc.edu; sxu@ias.edu}

\author{A. Lazarian}
	\affiliation{Department of Astronomy, University of Wisconsin, 475 North Charter Street, Madison, WI 53706, USA}
	\affiliation{Centro de Investigación en Astronomía, Universidad Bernardo O’Higgins, Santiago, General Gana 1760, 8370993, Chile}
	\author[0000-0002-0458-7828]{Siyao Xu\footnote{NASA Hubble Fellow}}
	\affiliation{Institute for Advanced Study, 1 Einstein Drive, Princeton, NJ 08540, USA}

\begin{abstract}
In a weakly compressible high-$\beta$ medium,
pitch-angle scattering and the associated scattering acceleration of cosmic rays (CRs) by anisotropic 
Alfv\'{e}n and slow modes of 
magnetohydrodynamic (MHD) turbulence
is inefficient. 
To tap the energy from magnetic compressions for efficient particle acceleration, 
a diffusion mechanism that can effectively confine particles in space without causing their trapping 
or pitch-angle isotropization
is needed. 
We find that 
the mirror diffusion in MHD turbulence recently identified in 
\citet{LXm21} 
satisfies all the above conditions and serves as a promising diffusion mechanism for efficient acceleration of CRs via their stochastic 
non-resonant interactions with magnetic compressions/expansions.
The resulting mirror acceleration is dominated by the slow-mode eddies with their lifetime comparable to the mirror diffusion time of CRs.
Consequently, we find that 
the acceleration time of mirror acceleration
is independent of the spatial diffusion coefficient of CRs. 
The mirror acceleration brings new life for the particle acceleration in a weakly compressible/incompressible medium
and has important implications for studying CR re-acceleration in the high-$\beta$ intracluster medium.


\end{abstract}

\section{Introduction} 

Cosmic rays (CRs) are the non-thermal energetic component of galaxies and galaxy clusters.
Studies on their diffusion and acceleration have far-reaching astrophysical implications on understanding the strong correlation between nonthermal emission and infrared luminosities of star-forming galaxies 
\citep{Yun01,Ajel20}
and influence of CRs on star formation, and galaxy and galaxy cluster evolution
(e.g., \citealt{Blas00,Pfro08,Hop12,Evoli12,La16,BruL16,Wien17,Hol18,Krum20,Seme21,Quat22,Liu23}.)

The interaction of CRs with the ubiquitous turbulent magnetic fields causes their diffusion and stochastic acceleration
\citep{Schlickeiser02,Mar20,Liu21}.
The advances achieved recently in theoretically understanding
\citep{GS95,LV99}
, simulating
(e.g., \citealt{MG01,CL02_PRL,Bere14}), 
and observationally measuring  
(e.g., \citealt{Laz18,Humc19,Huse22,Yuen22})
magnetohydrodynamic (MHD) turbulence 
bring significant changes to the standard paradigm of CR diffusion and acceleration
(e.g., \citealt{Chan00,YL04,XY13,LY14,XLb18,LucS18,Siou20,Hucr22,Kemp22,Bea22,LXm21,Forn21,Zhda21,Natt21,Lemo22,Samp23,Lemo23,Kemp23}).

In a weakly compressible high-$\beta$ medium, where $\beta$ is the ratio of the gas pressure to the magnetic pressure, both Alfv\'{e}n and slow modes of MHD turbulence have scale-dependent anisotropy 
\citep{CV00,CL02_PRL}
and are thus inefficient in scattering the CRs with the Larmor radii much smaller than the energy injection scale of turbulence 
\citep{Chan00,YL02,BYL2011,XL20}. 
Consequently, the spatial confinement and 
stochastic acceleration via gyroresonant scattering is inefficient. 
In a high-$\beta$ medium, 
the resonance-broadened Transit Time Damping
(TTD) by slow modes is found to dominate over that by fast mode in scattering and accelerating CRs 
\citep{XLb18}. 
However, as the ``pitch-angle scattering" by TTD is in fact caused by the stochastic acceleration in the direction parallel to the magnetic field, TTD alone cannot spatially confine CRs. 
It would lead to a highly anisotropic pitch angle distribution and cannot be self-sustained.

Non-resonant interactions of CRs with 
MHD turbulence have also been investigated for studying CR diffusion and stochastic acceleration
\citep{Noe68,CesK73,Ptuskin88,Klep95,CL06,Brunetti_Laz,Med15,BruL16,XZ17,Bres22,LXm21}.
The acceleration of CRs by magnetic compressions on scales larger than their Larmor radii was studied by 
\citet{CL06},
where
as the diffusion due to pitch-angle scattering was invoked, 
significant compression of gas is required for 
a net energy gain in average
(see also \citealt{Druf12}).
Therefore, their model cannot lead to efficient acceleration in a weakly compressible medium.

Based on the modern understanding of MHD turbulence, 
\citet{LXm21} recently identified the mirror diffusion of CRs parallel to the magnetic field. 
Unlike the mirror trapping 
with particles trapped between two magnetic mirror points, 
the mirror diffusion happens due to the perpendicular superdiffusion of turbulent magnetic fields 
\citep{LVC04,XY13,Bere13,Eyin13,LY14,Hucr22}.
CRs interact with magnetic compressions, i.e., magnetic mirrors, along the magnetic field lines
while experiencing superdiffusion in the direction perpendicular to the magnetic field. 
In a high-$\beta$ medium, as the pitch-angle scattering is inefficient, the mirror diffusion is expected to be the dominant mechanism for parallel diffusion
\citep{LXm21}.
\footnote{We note that other mechanisms in a high-$\beta$ medium can also contribute to confining CRs, e.g., field line wandering in super-Alfv\'{e}n turbulence 
\citep{Brunetti_Laz}, strong scattering by small-scale sharp magnetic field bends
\citep{Lemo23,Kemp23}.}
Unlike pitch-angle scattering, 
mirroring can spatially confine particles without causing stochastic change in their pitch angles. 
Therefore, it provides a promising diffusion mechanism for the non-resonant acceleration by large-scale\footnote{By ``large scale", we mean the scales larger than the particle Larmor radius.} magnetic compressions to be efficient in a weakly compressible medium.

In this work, by taking into account the mirror diffusion, we will reexamine the non-resonant acceleration of CRs by magnetic compressions in a high-$\beta$ medium, which we will term ``mirror acceleration".
In Section 2, we will review the mirror diffusion in a high-$\beta$ medium.
In Section 3, we will focus on the mirror acceleration in a high-$\beta$ medium. 
Conclusions are presented in Section 4.

\section{Mirror diffusion in a high-$\beta$ medium}

\subsection{Interaction of particles with magnetic mirrors}

For a particle with the total momentum $p$ moving along the magnetic field of strength $B$, the first adiabatic invariant is conserved during the mirroring process,
\begin{equation}
    J_1 \propto p^2\frac{1-\mu^2}{B},
    \label{J1}
\end{equation}
where the cosine of the pitch angle $\mu$ changes in accordance with the change of $B$, 
and pitch angle is the angle between the directions of ${\bm p}$ and ${\bm B}$. 

As the first condition 
for the mirroring to happen, 
the Larmor radius $r_g$ of the particle should be smaller than the size of the mirror, i.e., the variation scale of the magnetic field. 
As a particle gyrates moving along a magnetic field of increasing strength, its perpendicular
momentum $p_\perp = p \sqrt{1-\mu^2}$
increases (see Eq. \eqref{J1}). 
With a constant particle energy, it follows that 
the parallel momentum $p_\| = p \mu$
of the particle decreases. 
When the magnetic fluctuation is sufficiently large, the mirroring force is so strong that $p_\|$ can decrease to zero. 
Then the particle is reflected by the mirror. 
The condition for the mirror reflection is  
\begin{equation}
\frac{p_\bot^2}{B_0}=\frac{p^2}{B_0+\delta b},
\end{equation}
where $B_0$ and $B_0 + \delta b$ are the magnetic field strengths in the weak and strong magnetic field regions.
The maximum $\mu$ corresponding to the smallest pitch angle for mirror reflection is  
\begin{equation}
\mu^2_{max}=\frac{\delta b}{B_0+\delta b}.
\label{mu}
\end{equation}
The particles with $\mu < \mu_{max}$ are subject to mirroring. 


As the third condition for mirroring, 
the pitch-angle scattering rate cannot exceed the mirroring rate for $J_1$ to be conserved. 
In a low-$\beta$ medium, with efficient scattering by fast modes, mirroring happens over a range of $\mu$ with the mirroring rate larger than the scattering rate
\citep{CesK73,XL20,LXm21}.
For the weakly compressible high-$\beta$ medium considered in this work, 
fast modes are sound waves moving with an infinite velocity and their effect on particle dynamics is negligible. 
The scattering by Alfv\'{e}n and slow modes is inefficient at all $\mu$'s, 
and thus the mirroring occurs within the entire range 
$0<\mu<\mu_{max}$.

\subsection{Mirror diffusion of energetic particles}

\subsubsection{Mirror trapping and mirror diffusion}

{By following \citet{CesK73}}, 
in \citet{XL20}, the interaction of particles with MHD turbulence was 
separated into two regimes, {the classical diffusion caused by pitch-angle scattering} at $\mu>\mu_{c}$ and mirror trapping of particles at $\mu<\mu_{c}$. 
At the critical $\mu_c$, there is a balance between mirroring rate and scattering rate
\citep{XL20,LXm21}. 
{Within a limited range of pitch angles for scattering to dominate over mirroring, 
the corresponding scattering mean free path is reduced, and the problem of vanishing scattering close to $90^\circ$, i.e., $90^\circ$ problem
\citep{Jokipii1966}, 
is naturally resolved by mirroring. }

{In \citet{LXm21} (hereafter, LX21), the dynamics of mirroring particles at 
$\mu<\mu_{c}$ 
was reexamined. 
There it was found that due to the perpendicular superdiffusion of magnetic field lines in Alfv\'{e}nic turbulence
\citep{LV99,Eyin13,Laz20}, 
the mirroring particles can not be trapped between two mirroring points 
as they can not trace the same magnetic field line back and forth. 
Instead, they are stochastically reflected by different magnetic mirrors
while moving along the diverging turbulent magnetic field lines.
Thus a new diffusion process termed ``mirror diffusion" was identified. 
It is found to be much slower than the scattering diffusion
\citep{LXm21,Xu21}. }

\begin{figure}[htbp]
   \includegraphics[width=7cm]{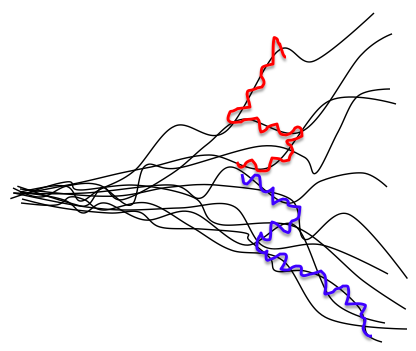}
\caption{ Illustration for mirror diffusion.
The perpendicular superdiffusion of {\it turbulent} magnetic field lines originally identified in \citet{LV99}, induces the parallel diffusion of particles that {stochastically interact with different} magnetic mirrors.  
Black lines are magnetic field lines. Blue and red lines are trajectories of two particles with a small initial separation.  From LX21. }
\label{fig: sket}
\end{figure}

Figure \ref{fig: sket} illustrates the process of mirror diffusion. The mirroring particles do not retrace their way back but follow new routes {led by stochastic magnetic field lines.} 
In a high-$\beta$ medium, the perpendicular superdiffusion induced by Alfv\'{e}n modes 
and the mirroring induced by slow modes together result in the mirror diffusion of CRs.

\subsection{Scattering diffusion and mirror diffusion}
\label{ssec:scatvsmirror}

Scattering diffusion is accompanied by the stochastic change of $\mu$. 
When a particle interacts with an eddy, 
if the scattering is so efficient that   
the timescale for pitch-angle isotropization is shorter than both the particle diffusion timescale 
and the lifetime of the eddy, 
the particle 
can simultaneously sample the compression (expansion) in the direction perpendicular to the magnetic field 
and the expansion (compression) in the direction parallel to the magnetic field, which accompany each other in a weakly compressible medium. 
After averaging over an isotropic distribution, 
energy gain and loss cancel out, and 
there is no net energy change. 
If the scattering is inefficient, 
particles freely move through the turbulent medium without being affected by magnetic compression/expansion, 
leading to inefficient acceleration. 

By contrast, 
mirror diffusion does not cause stochastic pitch angle change. 
A particle at a given $\mu$ can preferentially sample the compression/expansion in either perpendicular or parallel direction with respect to the magnetic field. 
Without averaging over an isotropic distribution, a net energy change
after interacting with an eddy is expected. 
Mirroring can also significantly suppresses the parallel diffusion of particles, which entails their sufficient interactions with magnetic compression/expansion. 
Therefore, mirror diffusion serves as a promising diffusion mechanism for efficient acceleration by large-scale magnetic compressions in a weakly compressible medium.

\subsubsection{Mirror diffusion induced by slow modes}
\label{f}

In a high-$\beta$ medium, 
slow modes of MHD turbulence create magnetic compressions that {serve as magnetic mirrors}. 
For simplicity, we assume that the slow mode fluctuations corresponding to the parallel wavenumber $k_{\|,sl}$, i.e. $\delta b=b_{sk}$, are small compared to the mean magnetic field strength $B_0$. Thus Eq. (\ref{mu}) is approximately 
\begin{equation}
\mu_{max}^2\approx \frac{b_{sk}}{B_0}.
\end{equation}
A particle with $\mu$ is most effectively reflected by the mirror 
with the magnetic fluctuation $b_{sk}$
corresponding to \citep{CesK73}
\begin{equation}
\mu^2\approx \frac{b_{sk}}{B_0}.
\label{mubk}
\end{equation}
The scaling of slow modes is
\citep{CL03} 
\begin{equation}
\label{bslow}
   b_{sk} \approx  \delta B_s (k_{\|, sl} L)^{-\frac{1}{2}},
\end{equation}
where $\delta B_s$ is the amplitude of slow mode fluctuations at the injection scale $L$. Therefore, combining Eqs. (\ref{mubk}) and (\ref{bslow}) leads to
\begin{equation}
k_{\|, sl} \approx \frac{1}{L} \left(\frac{\delta B_s}{B_0 \mu^2}\right)^2 =\frac{1}{L\mu^4} \aleph_s^{2},
\label{bkslow}
\end{equation}
where 
\begin{equation}
    \aleph_s =  \frac{\delta B_s}{B_0}.
\end{equation}
For weakly compressible MHD turbulence, the equivalent expression for $\aleph_s$ is
\begin{equation}\label{equns}
    \aleph_s \approx \frac{V_{L,slow}}{V_A},
\end{equation}
where $V_{L,slow}$ is {the turbulent velocity of slow modes} at $L$. With this notation, the critical cosine of pitch angle for mirroring is 
\begin{equation}
    \mu_{c}\approx \aleph_s^{1/2},
    \label{mualeph}
\end{equation}
which is the maximum $\mu$ determined by the largest magnetic fluctuation of slow modes. 
Particles with $\mu>\mu_c$ cannot be reflected through mirroring. 

The length corresponding to 
Eq. (\ref{bkslow})  
\begin{equation}
l_\mu \sim 1/k_{\|,sl},
\label{lmu}
\end{equation}
which acts as a proxy of the parallel mean free path of mirror diffusion at a given $\mu$ (LX21). 
{Its minimum value is determined by} 
\begin{equation}
 l_{\mu,min} = \text{max} [r_g, l_{\|,d}],
\end{equation}
where $l_{\|,d}$ is the parallel dissipation scale of slow modes. 
The particles with $\mu$ less than  
(Eq. \eqref{bkslow})
\begin{equation}
    \mu_{min}\approx \aleph_s^{1/2}\left(\frac{l_{\mu, min}}{L}\right)^{1/4}
\end{equation}
have the parallel mean free path equal to $l_{\mu,min}$.
Therefore, the parallel diffusion coefficient corresponding to mirror diffusion is 
\begin{equation}
    D_\| (\mu) \approx  v_p \mu l_\mu\approx v_p L \aleph_s^{-2} \mu^5  
    \label{Dpar}
\end{equation}
for $\mu > \mu_{min}$, 
where $v_p$ is the particle speed, and 
\begin{equation}
    D_\| (\mu) \approx  v_p \mu l_{\mu, min}
    \label{Dpar2}
\end{equation}
for $\mu < \mu_{min}$.

\section{Mirror acceleration by slow modes in a high-$\beta$ medium}

In this section, we disregard scattering of particles by slow and Alfv\'{e}n modes for simplicity, 
which are inefficient in 
scattering particles with 
$r_g \ll L$
due to the turbulence anisotropy
\citep{Chan00,YL02,XL20}. 
We consider only the second-order Fermi/stochastic acceleration of mirroing particles with $\mu<\mu_c$. The stochastic acceleration of scattering particles was analysed in \citet{CL06}.

\subsection{Electric field of slow modes}

{In addition to gyroresonant scattering, 
stochastic acceleration is frequently associated with 
non-resonant interaction of particles with large-scale magnetic compressions
(see \citealt{CL06}).} 
 The latter can occur in both compressible and incompressible media. 
In what follows, we will discuss the {non-resonant mirror acceleration} by slow modes in a weakly compressible medium. 
 
 In a weakly compressible media, the compression in one direction entails the expansion in {another} direction to 
 preserve the constant volume. 
{Accordingly, we will separately treat the perpendicular and parallel components of CR momentum.}

\begin{figure}
\centering
\includegraphics[width=0.9\linewidth]{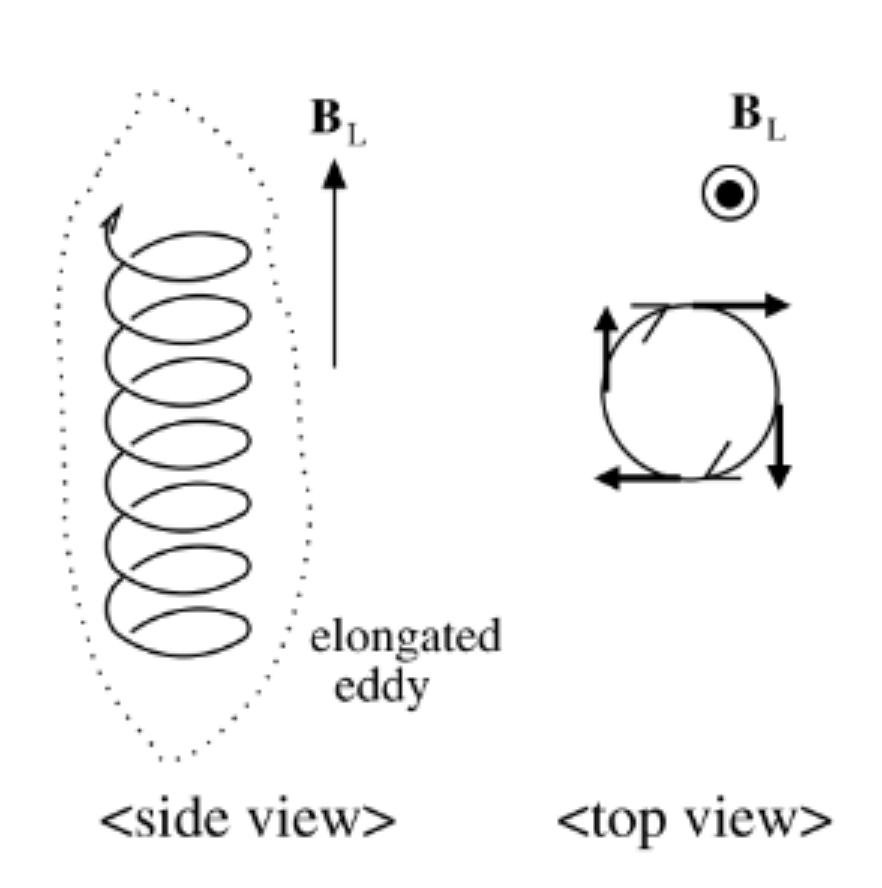}
\caption{The slow mode eddy is elongated along the local magnetic field direction, with a CR particle spiraling about the magnetic field and the electric field induced by the compression/expansion of the magnetic field. 
Modified from \citet{CL06}.}
\label{acceler}
\end{figure}

 Here we follow the approach  in 
\citet{CL06}. 
 Figure \ref{acceler} 
 illustrates the compression/expansion of the magnetic field and the acceleration/deceleration of a particle in a slow mode eddy. 
 The change of the magnetic field strength induces the electric field that is perpendicular to the magnetic field following the Faraday's law, 
 \begin{equation}
     E=-\frac{1}{2\pi r_g c}\int ({\bf B}_0 \cdot \bigtriangledown v_l) d{\bf s},
     \label{E1}
 \end{equation}
where $d{\bf s}$ is the infinitesimal area element vector, 
and $c$ is the light speed.
For longitudinal slow modes, 
the velocity perturbation along the magnetic field induces magnetic compression/expansion perpendicular to the magnetic field. 
So the spatial derivative of velocity in Eq. (\ref{E1}) can be approximated by $v_{l,\|}/l_{\|}$, 
where {$v_{l,\|}$ is the parallel turbulent velocity at the parallel length scale $l_\|$}. 
It represents the rate of compression or expansion of the magnetic field. 
As a result, taking the integral over the particle orbit area $\pi r_g^2$, it is easy to obtain
\begin{equation}
    E\approx -\frac{B_0 r_g v_{l,\|}}{2l_\| c} 
\end{equation}
By substituting the expression of $r_g=p_\bot c/(q B_0)$, where $q$ is the electric charge, we can rewrite the above equation as  
 \begin{equation}
E\approx -\frac{v_{l, \|} p_\bot}{2 q l_{\|}}.
\end{equation}

The electric field that a {particle} experiences results in the change of its $p_\perp$:
\begin{equation}
 \frac{dp_\bot}{dt} =q E \approx -\frac{p_\bot v_{l,\|}}{2l_{\|}},
\label{pper}
\end{equation}
{where we 
adopt 
{a negative sign for $v_{l,\|}/l_\|$ corresponding to magnetic compression 
and a positive sign corresponding to magnetic expansion.}. 
Naturally, strengthening of the magnetic field causes increase of $p_\perp$ based on the conservation of the first adiabatic invariant. 

\subsection{Gain and loss of energy} 



When the magnetic field is compressed in the perpendicular direction, according to Eq. (\ref{pper}), 
we have 
\begin{equation}\label{eq:persqr}
\frac{ \dot{p^2_\bot}}{p^2_\bot}=\frac{2\dot{p_{\bot}}}{p_{\bot}}=-\frac{v_{l,\|}}{l_\|},
\end{equation}
where $\dot{...}$ denotes the time derivative.
Compression of
the magnetic field entails increase of the squared perpendicular momentum,
which corresponds to energy gain. 
In an incompressible medium, compression in the perpendicular direction {with respect to the magnetic field} is inevitably accompanied by the expansion in the parallel direction. 
In the parallel direction,  the longitudinal invariant of a mirroring particle is 
\begin{equation}
J_2=\int^b_a p_{\|} d l_\|,
\end{equation}
where  
$a$ and $b$ are the two turning points.
Therefore, the expansion in the parallel direction results in the loss of energy, with
\begin{equation}
\frac{\dot{p^2_\|}}{p^2_\|}=\frac{2\dot{p_{\|}}}{p_{\|}}=2\frac{v_{l,\|}}{l_\|}.
\label{E2}
\end{equation}
By contrast, when the magnetic field is expanded in the perpendicular direction, 
energy loss occurs in the perpendicular direction, while energy gain occurs in the parallel direction.

 Given Eq. (\ref{eq:persqr}) and Eq.(\ref{E2}), the change of the squared total momentum of a {particle}, i.e.,  $p^2=p_\|^2+p_\bot^2$, {under the compression and expansion in different directions} is 
 \begin{equation}
 \dot{p^2}=-(p^2_\bot-2p^2_\|)\frac{v_{l,\|}}{l_{\|}}.
 \label{Etot}
 \end{equation}
As $p_\|=p \mu$ and $p_\bot=p \sqrt{1-\mu^2}$, Eq. (\ref{Etot}) can be rewritten as 
\begin{equation}
\frac{d p^2}{dt}=-p^2(1-3\mu^2) \left(\frac{v_{l,\|}}{l_\|}\right).
\label{p2}
\end{equation}
It shows that when encountering an eddy, 
the total energy gain or loss of a particle depends not only on the perpendicular compression/expansion of the magnetic field, but also on 
the sign of 
$(1-3\mu^2)$.

We rewrite {Eq. (\ref{p2})} to have the evolution of the total CR momentum,
\begin{equation}
\frac{d p}{dt}= -p \frac{1-3\mu^2}{2} \left(\frac{v_{l,\|}}{l_\|}\right).
\label{p1}
\end{equation}
{Formally}, the above equation is similar to Eq. (\ref{pper}) for $p_\perp$ apart from the 
{additional term $(1-3\mu^2)$.}
With $\mu<1/\sqrt{3} \approx 0.58$, 
the acceleration is dominated by that in the perpendicular direction.
The stochastic interaction of a particle with eddies with magnetic compression or expansion
induces the {stochastic} increase of $p$.
The acceleration tends to cause the decrease of $\mu$ and thus can be self-sustained with the mirroring condition always satisfied. 
With $\mu>1/\sqrt{3}$, 
the acceleration is attributed to the stochastic increase of 
$p_\|$.
With the increase of $\mu$, when $\mu>\mu_c$,
particles are not subject to mirroring, and the inefficient scattering alone cannot confine particles for them to be efficiently accelerated. 
Therefore, 
to ensure the mirror diffusion and mirror acceleration, 
in the following analysis, we consider 
the upper limit of $\mu$ as
\begin{equation}
    \mu_{ca}=min\Big[\mu_{c}, \frac{1}{\sqrt{3}}\Big],
    \label{MUC}
\end{equation}
where $\mu_c$ is given by Eq. \eqref{mualeph}.

We note that Eq. \eqref{p1} would lead to 
\begin{equation}
   \frac{dp}{dt} = 0 , 
\end{equation}
when it is averaged over an isotropic distribution with $\langle \mu^2 \rangle = 1/3$. 
As expected, efficient scattering cannot lead to acceleration
(see Section \ref{ssec:scatvsmirror}).
As scattering diffusion was adopted in 
\citet{CL06},
in their model 
significant gas compression is needed 
for acceleration to happen.

\subsection{Fast and slow parallel diffusion}

\subsubsection{Relevant timescales}

Next we will use Eq. (\ref{p1}) to evaluate the stochastic acceleration induced by {mirror} diffusion.


For a CR to {cross} an eddy of parallel size $l_\|$, the required time is
\begin{equation}
    \Delta t_d (\mu) \approx \frac{l_\|^2}{D_\| (\mu)},
    \label{td}
\end{equation}
where $D_\|(\mu)$ is $\mu$-dependent parallel diffusion coefficient of mirror diffusion given by Eqs. \eqref{Dpar} and \eqref{Dpar2}.
To define {different regimes of particle acceleration}, $\Delta t_d(\mu)$ should be compared with the {lifetime} of the slow mode eddy:
\begin{equation}
    \Delta t_e \approx \frac{l_\|}{V_A}.
    \label{te}
\end{equation}
Therefore, we can define 
the regime of {\it slow diffusion} with $\Delta t_d (\mu) > \Delta t_e$ 
and {\it fast diffusion} with $\Delta t_d (\mu)< \Delta t_e$. 
Apparently, both $\Delta t_e$ and $\Delta t_d (\mu)$ depend on the scale of turbulent eddies.
{Thus whether the diffusion is slow or fast 
depends on the range of length scales of interest.}

{The momentum diffusion results from the stochastic interaction of particles with eddies with magnetic compression/expansion.} To find the momentum diffusion coefficient
\begin{equation}
    D_p=\frac{(\Delta p)^2}{\Delta t},
    \label{D0}
\end{equation}
one needs to know the {change} of particle momentum $\Delta p$ {after interaction with an eddy of parallel size $l_\|$,}
\begin{equation}
    \Delta p\approx \frac{dp}{dt}\Delta t.
    \label{deltap}
\end{equation}
By substituting Eq. (\ref{p1}), it is easy to obtain
\begin{equation}
    \Delta p \approx - \frac{p_{eff} v_{l,\|}}{l_\|}\Delta t 
    \label{deltap1}
\end{equation}
where
\begin{equation}
    p_{eff}= p\frac{1-3\mu^2}{2} >0.
    \label{peff}
\end{equation}
We next derive $D_p$ in different diffusion regimes. 

\subsubsection{Fast parallel diffusion} 

The fast parallel diffusion (FPD) of particles over a scale $l_\|$ corresponds to the requirement:
\begin{equation}
    \Delta t_d (\mu) < \Delta t_e,
    \label{less0}
\end{equation}
which is equivalent to 
(Eqs. (\ref{td}) and (\ref{te})) 
\begin{equation}
    D_\| (\mu) > l_\| V_A.
    \label{geq}
\end{equation}
 As the time over which particles sample the magnetic compression/expansion within the eddy, 
 $\Delta t_d(\mu)$ enters
 Eq. (\ref{D0}), yielding 
\begin{equation}
    D_p^{FPD} (\mu) \approx \left(\frac{dp}{dt}\right)^2 \Delta t_d (\mu).
    \label{D0f}
\end{equation}
By further using Eqs. \eqref{td} and \eqref{deltap1}, 
we have 
\begin{equation}
    D_p^{FPD} (\mu)\approx \frac{v_{l,\|}^2 p_{eff}^2}{l_{\|}^2}\frac{l_{\|}^2}{D_\|(\mu)}=\frac{v_{l,\|}^2 p_{eff}^2}{ D_\| (\mu)},
    \label{Dpmu}
\end{equation}
where $D_\| (\mu)$ is given by Eqs. \eqref{Dpar} and \eqref{Dpar2}.

\subsubsection{Slow parallel diffusion}

 The case of {slow parallel diffusion (SPD)} is different.  For particles with
\begin{equation}
    \Delta t_d > \Delta t_e,
    \label{ne}
\end{equation}
i.e.,
\begin{equation}
    D_\| (\mu) < l_\| V_A,
    \label{leq}
\end{equation}
{$\Delta t_d (\mu)$ still enters $D_p$
in Eq. (\ref{D0}),
but $\Delta t_e$ enters $\Delta p$ in Eq. \eqref{deltap}.}
Thus we have 
\begin{equation}
\begin{aligned}
    D_p^{SPD} (\mu) &\approx \left(\frac{dp}{dt}\right)^2 \frac{(\Delta t_e)^2}{\Delta t_d (\mu)}  \\
  &  =\left[ \left(\frac{dp}{dt}\right)^2 \Delta t_d (\mu)\right]\left(\frac{\Delta t_e}{\Delta t_d (\mu)}\right)^2\\
  &  = D_p^{FPD}(\mu) \left(\frac{\Delta t_e}{\Delta t_d (\mu)}\right)^2.
    \label{Dslow0}
\end{aligned}
\end{equation}
As the ratio $\Delta t_e / \Delta t_d(\mu) <1$,
there is 
\begin{equation}
    D_p^{SPD} (\mu) < D_p^{FPD} (\mu).
\end{equation}
By using Eqs. \eqref{td}, \eqref{te}, and \eqref{Dpmu}, we have 
\begin{equation}
    D_p^{SPD} (\mu) \approx \left[ \frac{v_{l,\|}^2 p_{eff}^2}{ D_\| (\mu)}\right]\left(\frac{D_\|^2 (\mu)}{V_A^2 l_\|^2}\right).
    \label{Dslow1}
\end{equation}
According to Eq. (\ref{Dpar}), $D_{\|} (\mu)$ strongly depends on $\mu$. 
Thus we expect that the scales over which the parallel diffusion is fast or slow 
depend on $\mu$. 

\subsection{Acceleration for fast parallel diffusion}



For a given $\mu$, 
the FPD is applicable for scales smaller than a critical parallel scale $l_{\|,c}(\mu)$ corresponding to $\Delta t_d(\mu) = \Delta t_e$, 
which can be obtained by using Eqs. \eqref{Dpar}, \eqref{Dpar2}, 
\eqref{td}, and \eqref{te},
\begin{equation}
    l_{\|,c} (\mu)= \frac{D_{\|} (\mu)}{V_A} \approx L  \aleph_s^{-2} \left(\frac{v_p}{V_A}\right) \mu^{5}
    \label{lb}
\end{equation}
at $\mu_{min}<\mu<\mu_{ca}$, and 
\begin{equation}
   l_{\|,c}(\mu) \approx   l_{\mu,min} \left(\frac{v_p}{V_A}\right)\mu
\end{equation}
at $\mu<\mu_{min}$.
{For $l_{\|,c}(\mu)$ to be larger than $l_{\|,d}$, 
the above equations constrain 
\begin{equation}
   \mu > \mu_{min,FPD,1} = \left(\frac{l_{\|,d}}{L}\frac{V_A}{v_p} \aleph_s^{2}  \right)^\frac{1}{5}
\end{equation}
at $\mu_{min} < \mu < \mu_{ca}$, and 
\begin{equation}
   \mu> \mu_{min,FPD,2} = \frac{l_{\|,d}}{l_{\mu,min}}\frac{V_A}{v_p}
\end{equation}
at $\mu< \mu_{min}$.}
We see that when 
\begin{equation}
     v_p > \Big(\frac{l_{\|,d}}{l_{\mu,min}}\Big)^\frac{5}{4} \Big(\frac{l_{\|,d}}{L}\Big)^{-\frac{1}{4}}
     \aleph_s^{-\frac{1}{2}} V_A,
\end{equation}
there is $\mu_{min,FPD,1}>\mu_{min,FPD,2}$ and vice versa.
The minimum $\mu$ for the acceleration in the FPD regime is 
\begin{equation}
  \mu_{min,FPD} = \max[\mu_{min,FPD,1},\mu_{min,FPD,2}].
\end{equation}

For $l_{\|,d}<l_\|<l_{\|,c} (\mu)$, 
the momentum diffusion coefficient at a given $\mu$ is 
(Eqs. \eqref{equns} and \eqref{Dpmu})
    \begin{equation}
    D_p^{FPD} (\mu) 
    \approx \frac{V_A^2 p_{eff}^2 l_\|}{ D_\| (\mu)L} \aleph_s^2,
    \label{Dmufast}
\end{equation}
where the scaling of the slow modes (see Eq.(\ref{bslow})) 
\begin{equation}
    v_{l,\|}\approx V_{L,slow} \left(\frac{l_\|}{L}\right)^{1/2}
    \label{scalslow}
\end{equation}
is used.
The resulting  averaged momentum diffusion coefficient is 
\begin{equation}
\begin{aligned}
    \bar{D}_p^{FPD} (\mu)
    &= \int_{l_{\|, d}}^{l_{\|, c} (\mu)} D_p^{FPD} (\mu) \frac{dl_\|}{l_{\|}} \\
  & \approx \frac{V_A^2 p_{eff}^2}{D_\| (\mu) L} \aleph_s^2 \left[l_{\|,c}(\mu)-l_{\|,d}\right].
    \label{Dbarf}
\end{aligned}
\end{equation}
In the case with $\mu\gg \mu_{min,FPD}$ and the scale $l_{\|,c}(\mu)\gg l_{\|, d}$, we approximately have
\begin{equation}
    \bar{D}_p^{FPD} (\mu)\approx \frac{V_A p_{eff}^2}{ L} \aleph_s^2, 
    \label{Dbarf}
\end{equation}
where $l_{\|,c}(\mu) = D_\|(\mu)/V_A$ is used. 
{Its dependence on $\mu$ only comes from the dependence of $p_{eff}$ on $\mu$.}
We see that among the eddies of different parallel sizes, the ones with
the lifetime comparable to the particle diffusion time dominate the acceleration.


\subsection{Acceleration for slow parallel diffusion with $\mu \in (\mu_{min,FPD}, \mu_{ca})$}

Within the range $\mu \in (\mu_{min,FPD}, \mu_{ca})$, 
{the parallel diffusion over scales $l_\|>l_{\|,c} (\mu)$ falls in the SPD regime. 
The averaged momentum diffusion coefficient in the SPD regime is }
\begin{equation}
    \bar{D}_p^{SPD} (\mu)\approx \int_{l_{\|, c} (\mu)}^{L} D_p^{SPD} (\mu) \frac{dl_\|}{l_{\|}}.
\end{equation}
By using Eqs. (\ref{Dslow1}) and \eqref{scalslow}, we have 
\begin{equation}
    \bar{D}_p^{SPD} (\mu)\approx \frac{D_\| (\mu) p_{eff}^2}{L} \aleph_s^2 \left[\frac{1}{l_{\|, c}}-\frac{1}{L}\right].
\end{equation}
In the case with 
$l_{\| c}\ll L$, by using $l_{\|,c}(\mu) = D_\|(\mu)/V_A$, one can further write
\begin{equation}
    \bar{D}_p^{SPD} (\mu)\approx \frac{V_A p_{eff}^2}{ L} \aleph_s^2,
     \label{Dbars}
\end{equation}
which is the same as $\bar{D}_p^{FPD}(\mu)$ in Eq. \eqref{Dbarf}.
It shows that 
in our approximate treatment, { 
despite the opposite dependence of $D_p^{FPD}(\mu)$ (Eq. \eqref{Dpmu}) and $D_p^{SPD}(\mu)$ (Eq. \eqref{Dslow1}) on $D_\|(\mu)$, 
both $\bar{D}_p^{FPD}(\mu)$ and $\bar{D}_p^{SPD}(\mu)$ are dominated by the stochastic acceleration at $l_{\|,c}(\mu)$
and do not depend on $D_\|(\mu)$. 
The 
fast and slow parallel diffusion equally contribute to the momentum diffusion and acceleration.} 


\subsection{Acceleration for slow parallel diffusion with $\mu \in (0, \mu_{min,FPD})$}

For  $\mu \in (0, \mu_{min,FPD})$, the diffusion over all length scales $[l_{\|, d}, L]$ is always in the SPD regime, with 
\begin{equation}
  D_\| (\mu) < l_{\|,d} V_A=D_\|(\mu_{min,FPD}) .
  \label{eq:mumumin}
\end{equation}
{Given $l_{\|,d} \ll L$,} we have 
\begin{equation}
\begin{aligned}
    \bar{D}_p^{SPD} (\mu) &= \int_{l_{\|, d}}^{L} D_p^{SPD} (\mu) \frac{dl_\|}{l_{\|}}\\
    &\approx \frac{D_\| (\mu) p_{eff}^2}{L l_{\|,d}} \aleph_s^2\\
    &= \frac{V_A p_{eff}^2}{L} \aleph_s^2 \frac{D_\| (\mu)}{D_\| (\mu_{min,FPD})},
    \label{eq:Dmin}
\end{aligned}
\end{equation}
{which depends on $D_{\|}(\mu)$.}
As $D_\| (\mu) < D_\| (\mu_{min,FPD})$,
{$\bar{D}_p^{SPD}(\mu)$ with $\mu<\mu_{min,FPD}$ is smaller than that at a larger $\mu$ (Eq. \eqref{Dbars}).}
It indicates that 
the acceleration of particles with $\mu<\mu_{min,FPD}$ is less efficient than that with $\mu>\mu_{min,FPD}$.


\subsection{Total momentum diffusion coefficient}

{The averaged total momentum diffusion coefficient $\bar{D}_p^{total}(\mu)$ is given by 
$\bar{D}_p^{SPD}(\mu)$ in Eq. \eqref{eq:Dmin} at $\mu< \mu_{min,FPD}$.}
For $\mu \in (\mu_{min,FPD}, \mu_{ca})$, 
the SPD and FPD processes should be both taken into account, leading to 
\begin{equation}\label{eq:totpu}
    \bar{D}_p^{total} (\mu)\approx D_p^{FPD} (\mu) + D_p^{SPD} (\mu)\approx \frac{2V_A p_{eff}^2}{ L} \aleph_s^2,
\end{equation}
where we sum up the results in Eq. (\ref{Dbarf}) and Eq. (\ref{Dbars}).

{Furthermore, by averaging over pitch angles, we have} 
\begin{equation}
\begin{aligned}
    \bar{D}_p^{total}
    &=\frac{1}{\int_0^{\pi/2}f d\theta}
    \int_{\theta_{ca}}^{\pi/2} \bar{D}^{total}_p (\mu(\theta)) f (\mu(\theta)) d\theta \\
    &=\frac{1}{\int_0^{\pi/2}f d\theta} \int_0^{\mu_{ca}} \bar{D}_p^{total} (\mu) f (\mu) \frac{d\mu}{\sqrt{1-\mu^2}}.
    \label{Dpfast}
\end{aligned}
\end{equation}
For simplicity, in the case with a uniform pitch angle distribution, i.e., $f=1$,
\footnote{Note that the uniform pitch angle distribution considered here is different from the fast pitch-angle isotropization achieved by efficient pitch-angle scattering
during the interaction of a particle with an eddy discussed in Section \ref{ssec:scatvsmirror}.}
there is 
(Eqs. \eqref{peff} and \eqref{eq:totpu})
\begin{equation}\label{eq:dptamca}
\begin{aligned}
    \bar{D}_p^{total}
    &\approx \frac{1}{\pi} \frac{V_A p^2}{L} \aleph_s^2 \int_{\mu_{min,FPD}}^{\mu_{ca}} (1-3\mu^2)^2 \frac{d\mu}{\sqrt{1-\mu^2}} \\
    &\approx \frac{1}{\pi} \frac{V_A p^2}{L} \aleph_s^2 \int_{\mu_{min,FPD}}^{\mu_{ca}} (1-3\mu^2)^2 d\mu \\
    & \approx  \frac{1}{\pi} \frac{V_A p^2}{L} \aleph_s^2 \left(\frac{9\mu_{ca}^5}{5}-2\mu_{ca}^3+\mu_{ca}\right)\\
    & \approx \frac{1}{\pi} \frac{V_A p^2}{L} \aleph_s^2 \mu_{ca},
\end{aligned}    
\end{equation}
{where we assume that $\mu_{min,FPD}$ is close to $0$ and consider $\mu_{ca}\leq 1/\sqrt{3}$.}
The corresponding acceleration timescale can be estimated as  
\begin{equation}\label{tauacc}
\begin{aligned}
    \tau_{acc} &\approx \frac{p^2}{\bar{D}_p^{total}} \\
    &\approx \pi \frac{L}{V_A}  \aleph_s^{-2} 
    \left(\frac{9\mu_{ca}^5}{5}-2\mu_{ca}^3+\mu_{ca}\right)^{-1}\\
    & \approx \pi \frac{L}{V_A} \aleph_s^{-2} \mu_{ca}^{-1}.
\end{aligned}
\end{equation}
{We see that both $\bar{D}_p^{total}$ in Eq. \eqref{eq:dptamca} and 
$\tau_{acc}$ in Eq. \eqref{tauacc} are independent of $D_{\|}(\mu)$.}
This is due to the fact that 
irrespective of $D_\|(\mu)$, it is the eddies with their lifetime equal to the diffusion time corresponding to $D_\|(\mu)$ that dominate the acceleration.

\subsection{Pitch angle diffusion due to mirror acceleration}

The mirror acceleration results in {change of $p$ and thus change of $\mu$.} 
Similar to the transit time damping (TTD), the change of $\mu$ is associated with acceleration.
But unlike the resonant TTD that causes the stochastic increase of $\mu$, 
the non-resonant mirror acceleration causes the stochastic decrease of $\mu$.

The evolution of $\mu$ can be determined by using Eq. (\ref{pper}),
\begin{equation}
    \frac{d}{dt} (p\sqrt{1-\mu^2}) \approx -\frac{1}{2}p\sqrt{1-\mu^2} \frac{v_{l,\|}}{l_\|},
\end{equation}
which provides
\begin{equation}
    \sqrt{1-\mu^2} \frac{dp}{dt} - p \frac{\mu}{\sqrt{1-\mu^2}} \frac{d\mu}{dt}\approx -\frac{1}{2}p\sqrt{1-\mu^2} \frac{v_{l,\|}}{l_\|}.
    \label{eq:deriv}
\end{equation}
The substitution of Eq. (\ref{p1}) in Eq. (\ref{eq:deriv}) provides
\begin{equation}\label{dmu2new}
    \frac{d\mu^2}{dt} =3\mu^2 (1-\mu^2)  \frac{v_{l,\|}}{l_\|}.
\end{equation}
Then the time derivative $d\mu/dt$ is 
\begin{equation}\label{eq:dmudt}
\frac{d\mu}{dt} = \frac{3}{2} \mu (1-\mu^2) \frac{v_{l,\|}}{l_\|}.
\end{equation}

{Similar to the derivation of momentum diffusion coefficient, the derivation of pitch angle diffusion coefficient also depends on the diffusion regime.}
In the FPD regime, the time interval relevant in determining $\Delta \mu$ is $\Delta t_d(\mu)$, i.e.,
$\Delta \mu \approx (d\mu/dt) \Delta t_d(\mu)$. The time step for the diffusion is also $\Delta t_d(\mu)$. Thus the pitch angle diffusion coefficient in the FPD regime is 
(Eqs. \eqref{equns}, \eqref{scalslow}, \eqref{td}, and \eqref{eq:dmudt})
\begin{equation}
\begin{aligned}
    D_{\mu\mu}^{FPD} (\mu)
    &\approx \frac{(\Delta \mu)^2 }{\Delta t} \\
    &\approx \left(\frac{d\mu}{dt}\right)^2 \Delta t_d(\mu) \\
    & = \frac{9 \mu^2(1-\mu^2)^2 V_A^2 l_\|}{4 D_\|(\mu)L}  \aleph_s^2.
    \label{Dmm}
\end{aligned}
\end{equation}
Its dependence on $\mu$ is explicitly shown.
At $\mu_{min,FPD} <\mu<\mu_{ca}$, the averaged pitch angle diffusion coefficient is 
\begin{equation}
\begin{aligned}
   \bar{D}_{\mu\mu}^{FPD}(\mu) 
   &= \int_{l_{\|,d}}^{l_{\|,c}(\mu)}D_{\mu\mu}^{FPD}(\mu) \frac{dl_\|}{l_\|} \\ 
   &\approx  \frac{9 \mu^2(1-\mu^2)^2 V_A^2}{4 D_\|(\mu)L}  \aleph_s^2 [l_{\|,c}(\mu)-l_{\|,d}].
\end{aligned}
\end{equation}
Provided $l_{\|,c}(\mu)\gg l_{\|,d}$ and using $D_{\|}(\mu) = V_A l_{\|,c}(\mu)$, we approximately have 
\begin{equation}\label{eq:appdmmf}
\bar{D}_{\mu\mu}^{FPD}(\mu) \approx 
  \frac{9 \mu^2(1-\mu^2)^2 V_A}{4 L}  \aleph_s^2 .
\end{equation}
{It does not depend on $D_\|(\mu)$.}

In the SPD regime, 
we have $\Delta \mu\approx (d\mu/dt) \Delta t_e$. 
As a result, there is 
\begin{equation}
\begin{aligned}
    D_{\mu\mu}^{SPD}(\mu)
    &\approx \left( \frac{d\mu}{dt}\right)^2 \frac{(\Delta t_e)^2}{\Delta t_d(\mu)} \\
    &= \left[\left(\frac{d\mu}{dt}\right)^2 \Delta t_d(\mu)\right] \left(\frac{\Delta t_e}{\Delta t_d(\mu)}\right)^2 \\
    & \approx D_{\mu\mu}^{FPD}(\mu) \left(\frac{\Delta t_e}{\Delta t_d(\mu)}\right)^2 \\
    & \approx 
     \frac{9 \mu^2(1-\mu^2)^2 V_A^2 l_\|}{4 D_\|(\mu)L}  \aleph_s^2 \left(\frac{D_{\|}^2(\mu)}{V_A^2 l_\|^2}\right),
\end{aligned}
\end{equation}
where Eqs. \eqref{td}, \eqref{te}, and \eqref{Dmm} are used. 
With $D_\|(\mu) < V_A l_\|$, we naturally have 
\begin{equation}
   D_{\mu\mu}^{SPD}(\mu) < D_{\mu\mu}^{FPD}(\mu).
\end{equation}
At $\mu_{min,FPD}<\mu<\mu_{ca}$, the averaged pitch angle diffusion coefficient in the SPD regime is 
\begin{equation}
\begin{aligned}
  \bar{D}_{\mu\mu}^{SPD}(\mu)
  &= \int_{l_{\|,c}(\mu)}^L D_{\mu\mu}^{SPD}(\mu) \frac{dl_\|}{l_\|} \\
  &\approx \frac{9\mu^2(1-\mu^2)^2D_\|(\mu)}{4L}\aleph_s^2\left(\frac{1}{l_{\|,c}(\mu)}-\frac{1}{L}\right).
\end{aligned}
\end{equation}
Given $l_{\|,c}(\mu)\ll L$ and using the relation $D_{\|}(\mu) = V_A l_{\|,c}(\mu)$, the above expression is approximately
\begin{equation}\label{eq:apdmmsbm}
\bar{D}_{\mu\mu}^{SPD}(\mu) \approx
\frac{9\mu^2(1-\mu^2)^2 V_A}{4L}\aleph_s^2,
\end{equation}
which is the same as 
$\bar{D}_{\mu\mu}^{FPD}(\mu)$ in Eq. \eqref{eq:appdmmf}.

At $\mu< \mu_{min,FPD}$, the averaged pitch angle diffusion coefficient is 
\begin{equation}
\begin{aligned}
  \bar{D}_{\mu\mu}^{SPD}(\mu) &=
  \int_{l_{\|,d}}^{L} D_{\mu\mu}^{SPD}(\mu) \frac{dl_\|}{l_\|} \\
&\approx   \frac{9\mu^2(1-\mu^2)^2D_\|(\mu)}{4L}\aleph_s^2\left(\frac{1}{l_{\|,d}}-\frac{1}{L}\right).
\end{aligned}
\end{equation}
Given $l_{\|,d}\ll L$, it approximately becomes 
\begin{equation}
\bar{D}_{\mu\mu}^{SPD}(\mu) \approx 
   \frac{9\mu^2(1-\mu^2)^2 V_A }{4L}\aleph_s^2\frac{D_\|(\mu)}{D_\|(\mu_{min,FPD})},
\end{equation}
where Eq. \eqref{eq:mumumin} is used. It depends on $D_{\|}(\mu)$. 
With $D_{\|}(\mu) < D_{\|}(\mu_{min,FPD})$, 
$\bar{D}_{\mu\mu}^{SPD}(\mu)$ at $\mu<\mu_{min,FPD}$ is smaller than that at a larger $\mu$ (Eq. \eqref{eq:apdmmsbm}). 
{The more efficient pitch angle diffusion at $\mu > \mu_{min,FPD}$ is caused by the more efficient acceleration. }

\section{Conclusions}

In a weakly compressible medium, 
both mirror trapping and pitch-angle scattering cannot lead to acceleration of CRs by magnetic compressions. 
In the former case with a 
particle trapped within a single magnetic bottle with 
oscillating magnetic compression and expansion and conserved adiabatic invariants, the stochastic acceleration does not happen.
In the latter case with efficient scattering, in each eddy a particle 
undergoes fast pitch-angle isotropization 
and 
samples both fluid compression and expansion 
that happen simultaneously in different directions, and thus its energy gain and loss cancel out. 
Moreover, in realistic MHD turbulence, 
mirror trapping does not happen, and the 
scattering by anisotropic Alfv\'{e}n and slow modes are inefficient.

As a new diffusion mechanism,  
the mirror diffusion in MHD turbulence
identified by LX21
does not cause trapping of particles or pitch-angle isotropization. 
It takes place due to the perpendicular superdiffusion of turbulent magnetic fields regulated by Alfv\'{e}n modes 
and mirroring by slow modes. It can effectively confine CRs via stochastic mirror reflection in the direction parallel to the magnetic field. 
Under the consideration of mirror diffusion, 
the stochastic non-resonant interaction with slow-mode eddies with magnetic compression/expansion results in the mirror acceleration. 
The mirror acceleration takes place 
irrespective of the compressibility of gas.

Among the slow-mode eddies with different parallel sizes along the turbulent energy cascade, we find that the ones with their lifetime comparable to the mirror diffusion timescale dominate the mirror acceleration. 
It follows that the resulting momentum diffusion coefficient does not depend on the mirror diffusion coefficient. 
The acceleration time only depends on the Alfv\'{e}n crossing time and relative magnetic fluctuation of slow modes at the driving scale of turbulence. 
In comparison with the inefficient scattering acceleration associated with inefficient scattering diffusion 
\citep{CL06}, 
the mirror acceleration serves as an efficient acceleration mechanism in a high-$\beta$ medium.

The mirror acceleration causes stochastic increase of CR perpendicular momentum and pitch angle, and thus the acceleration can be self-sustained with the condition for mirroring satisfied. 
The mirror acceleration in a weakly compressible high-$\beta$ medium
can be applied to studying CR re-acceleration in the high-$\beta$ intracluster medium. 
This application will be investigated in our future study.

    

\acknowledgments
 A.L. acknowledges the support of the NSF grants AST1715754, 1816234 and NASA ATP AAH7546.   The  Flatiron  Institute  is supported by the Simons Foundation.
S.X. acknowledges the support for 
this work provided by NASA through the NASA Hubble Fellowship grant \# HST-HF2-51473.001-A awarded by the Space Telescope Science Institute, which is operated by the Association of Universities for Research in Astronomy, Incorporated, under NASA contract NAS5-26555. 
S.X. also acknowledges the support from the Institute for Advanced Study.

\bibliographystyle{aasjournal}
\bibliography{xu}

\end{document}